# Asymmetric nonlinear optics of a polar chemical bond

Yuya Morimoto[1,2*†], Yasushi Shinohara[3,4*], Mizuki Tani[4], Bo-Han Chen[1,2],
Kenichi L. Ishikawa[3,4,5], and Peter Baum[1,2,6]

[1] *Ludwig-Maximilians-Universität München, Am Coulombwall 1, 85748 Garching, Germany*

[2] *Max-Planck-Institute of Quantum Optics, Hans-Kopfermann-Str. 1, 85748 Garching, Germany*

[3] *Photon Science Center, Graduate School of Engineering, The University of Tokyo, 7-3-1 Hongo, Bunkyo-ku, Tokyo 113-8656, Japan*

[4] *Department of Nuclear Engineering and Management, Graduate School of Engineering, The University of Tokyo, 7-3-1 Hongo, Bunkyo-ku, Tokyo 113-8656, Japan*

[5] *Research Institute for Photon Science and Laser Technology, The University of Tokyo, 7-3-1 Hongo, Bunkyo-ku, Tokyo 113-0033, Japan*

[6] *University of Konstanz, Universitätsstraße 10, 78457 Konstanz, Germany*

Corresponding author. Email: yuya.morimoto@fau.de (Y.M.), ishiken@n.t.u-tokyo.ac.jp (K.L.I), and peter.baum@uni-konstanz.de (P.B.)

*These authors contributed equally to this work.

[†] Current address: *Friedrich-Alexander-Universität Erlangen-Nürnberg, Staudtstraße 1, 91058, Erlangen*

Dated: October 25, 2019

**Abstract**

A dielectric material's response to light is macroscopically described by electric displacement fields due to polarization and susceptibility, but the atomistic origin is light-cycle-driven motion of electron densities in the restoring forces of the atomic environment. Here we report how the macroscopic nonlinear-optical response of a heteronuclear crystal relates to the alignment and orientation of its chemical bonds. Substantial nonlinear emission is only observed if the electric field of an optical single-cycle pulse points from the less electronegative to the more electronegative element and not vice versa. This asymmetry is a consequence of the unbalanced real-space motion of valence charges along the direction of the bonds. These results connect a material's chemical structure to the optical response and may facilitate the comprehension and design of novel materials for applications in optics and lasers on basis of the atoms and how they connect.

**Main text**

When a light wave interacts with a transparent material, the electric field of the optical cycles produce a time-dependent polarization and electric displacement fields via the material's linear or nonlinear response to the optical cycles (*1, 2*). The light field that is emitted is therefore often complicated and gives rise to a wealth of optical phenomena with fundamental and technical relevance, for example the refractive index, birefringence, dispersion, optical parametric amplification, frequency-doubling, self-phase modulation, multi-photon absorption, higher-harmonic generation and many more. Textbooks treat perturbative linear and nonlinear optics via a time-dependent polarization density $P(t) = \epsilon_0[\chi^{(1)}E(t) + \chi^{(2)}E^2(t) + \chi^{(3)}E^3(t) + \cdots]$, where $\chi^{(n)}$ is the *n*-th order susceptibility tensor and $E(t)$ is the applied electric field. In the non-perturbative and strong-field regimes, *P(t)* depends on the cycle-level history of *E(t)* in time, for example via field-induced intraband currents (*3, 4*), dynamical metallization (*5*), core excitons (*6*), sub-cycle energy transfer (*7*), dynamical electron localization to *d*-orbitals (*8*), or seed-induced changes of the band structure (*9*). Although many such nonlinear-optical and strong-field phenomena in atoms, molecules, solids and nanostructures are currently rather well understood on basis of fields and polarizations, we lack a comprehensive understanding of the atomistic origin of the macroscopic quantities *P(t)* and $\chi^{(n)}$ and how the structure of a material causes the effective nonlinear-optical phenomena that we observe. Early attempts to link $\chi$ and *P* to a material's atomic structure include the Clausius-Mossotti relation and the Lorentz–Lorenz equation which predict the dielectric constant and refractive index but not any nonlinear effects. Modern experiments in which amorphous or crystalline materials are excited by intense, short laser pulses (*3, 10–20*) usually invoke dynamical quantum-calculations on a band structure (*21*) for explaining the results, and a real-space, atomistic explanation is often not attempted, although it could provide a connection between the crystal structure and the nonlinear optical response (*16, 22, 23*).

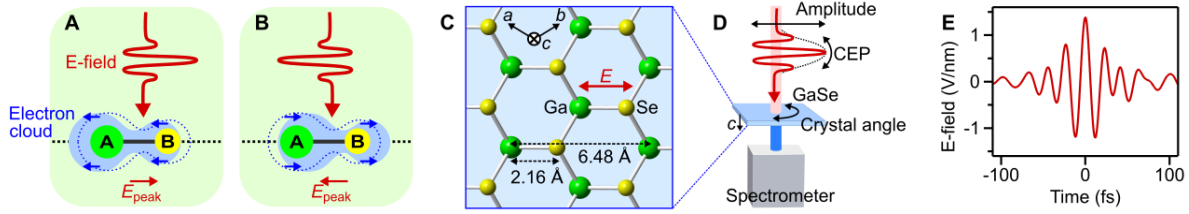

**Fig. 1. Concept and experiment. (A)**, **(B)** An asymmetric chemical bond (solid line) in a crystal is excited by a single-cycle laser pulse (red). Depending on the direction of the most intense electric field cycle ($E_{peak}$), electron density (blue) is coherently driven along or opposite to the polarity of the bond (atoms A and B). The field-driven electrons emit new light with a waveform determined by the atoms' electronegativity. **(C)** Crystal structure of one monolayer of $\varepsilon$-GaSe crystal in the *a-b* plane. **(D)** Experiment. Near-single-cycle mid-infrared laser pulse (red) is focused into a $\varepsilon$-GaSe crystal (blue) and the output spectrum is analysed as a function of crystal angle and peak field direction. **(E)** Electric field of the driving pulse in the experiment.

Here we excite a single-crystal of $\varepsilon$-GaSe and its heterogeneous chemical bonds with near-single-cycle pulses of mid-infrared light and relate the nonlinear optical response to the direction and polarity of the bonds. The excitation consists of only one relevant optical cycle (1.6 field cycles in the pulse width) and therefore approximately explores the optical impulse response in time. We find a pronounced asymmetry of the nonlinearity if the excitation's electric field points from a Ga atom to a Se atom as compared to the other direction. The two simplest possible models, a point charge in an effective asymmetric crystal potential and a one-dimensional quantum simulation with two different atomic potential wells deliver both a good agreement, showing that the strengths and orientations of the chemical bonds control and determine the optical response, in particular the timing structure and absolute direction of the nonlinear polarization.

The experiment is depicted in Fig. 1D. A near-single-cycle optical pulse (1.6 cycle FWHM, see Fig. 1E) with a central frequency of 43 THz (7 μm wavelength) is obtained via optical parametric amplification (*24*) and focused into a 21-μm thick single-crystal of gallium selenide ($\varepsilon$-GaSe). The optical spectrum does not excite any infrared active phonon modes (*25, 26*) and the photon energy of ~0.18 eV is far below the band gap of 2.0 eV (*27*). The peak field strength inside the crystal is 1.4 V/nm and the electric field is parallel to the Ga-Se bonds in the *a-b* plane (Fig. 1C) (*23, 26*). The output spectrum after the nonlinear interaction is collected with a silver-coated spherical mirror (2-inch focal length) and guided via an InF$_3$ multimode fiber (Le Verre Fluoré) into threse types of spectrometers, namely a Fourier transform spectrometer (L-FTS, LASNIX) for the mid-infrared range, a grating-based spectrometer with a InGaAs detector for the near-infrared range (Rock NIR RSM-445, Ibsen Photonics) and a spectrograph with a silicon detector for visible light (USB-2000+, ocean optics). These spectrometers have calibrated sensitivities and overlapping spectral ranges that allow to determine a concatenated result (*14*).

Figure 2A shows the observed output spectra from the GaSe crystal as a function of the excitation pulse's carrier-envelope phase (CEP, $\varphi_{CE}$). We see a broadband nonlinear light emission that ranges from the fundamental frequency (43 THz, λ=7.0 µm) to a more than 17-fold higher frequency of more than 750 THz (λ= 0.4 µm); the optical fiber absorbs the light beyond that range. Depending on the CEP, the spectrum consists of a pronounced sequence of harmonic orders ($\varphi_{CE} = \pi = 180°$) or alternatively covers smoothly the whole range without interferences (*3*, *18*, *20*) . A common peak around 480 THz is incoherent photoluminescence (PL) at the band gap of ~2 eV from residual nonlinear absorption. In Figs. 2B and 2C, we show the spectral details at two special CEP values, $\varphi_{CE} = 0$ and $\varphi_{CE} = \pi = 180°$, the white lines in Fig. 2A. At these values, the temporal excitation waveforms are identical and have a single optical cycle that is most intense, but the absolute direction of the electric field is flipped. We call $\varphi_{CE} = 0$ a cosine excitation and $\varphi_{CE} = \pi$ a minus-cosine excitation. At $\varphi_{CE} = 0$, the spectrum is smooth, while at $\varphi_{CE} = \pi$ the spectrum shows a set of distinct harmonic peaks at a visibility of more than 100. There are even-order and odd-order harmonics with a spacing given by one photon energy of the driving frequency (43 THz, 0.18 eV).

In order to understand the physics behind our observations, we recall the crystal structure of *ε*-GaSe as shown in Figs. 1C and S2. The crystal is composed of quasi-2D monolayers that are bonded to each other with van der Waals forces. Our driving electric field oscillates in the *a-b* plane (see Fig. 1C) and inter-layer electron dynamics along the *c*-axis can therefore be neglected (*23*). An electric pump field vector in the *a-b*-plane will predominantly project only onto one of the three available Ga-Se bonds (see Figs. 1C and S2) while the other two bonds at 60-degree bond angles in the *a-b* plane will see an effectively much weaker projection of the electric field (*26*). The absolute orientation of all Ga-Se bonds is hence unique in *ε*-GaSe throughout all of the unit cell and there are no opposite Se-Ga bonds excited. We therefore concentrate our analysis on the real-space dynamics of a single chemical bond.

We first consider the simplest possible model and assume that the dynamics of valence electrons is represented by a classical point charge that moves one-dimensionally along the bond direction in the restoring forces of an asymmetric atomic potential; see Fig. 2D. We describe the model potential $U(x)$ by a series of polynomials, $U(x) = \sum_{n=2}^{\infty} c_n x^n$, where $c_2$ gives the linear-optical response and $c_n$ ($n > 2$) describe the nonlinearity in the motion of the charge; the odd-order terms determine the asymmetry of the potential. We determine $c_2$ from the direct band gap (*26*) and $c_3 - c_6$ from the harmonic intensities in Fig. 2C; higher orders are neglected (*26*). The measured electric field of the driving field (see Fig. 1E) is used to compute the real-space motion of the charge and Fourier transformation of the acceleration gives the output spectrum (*26*).

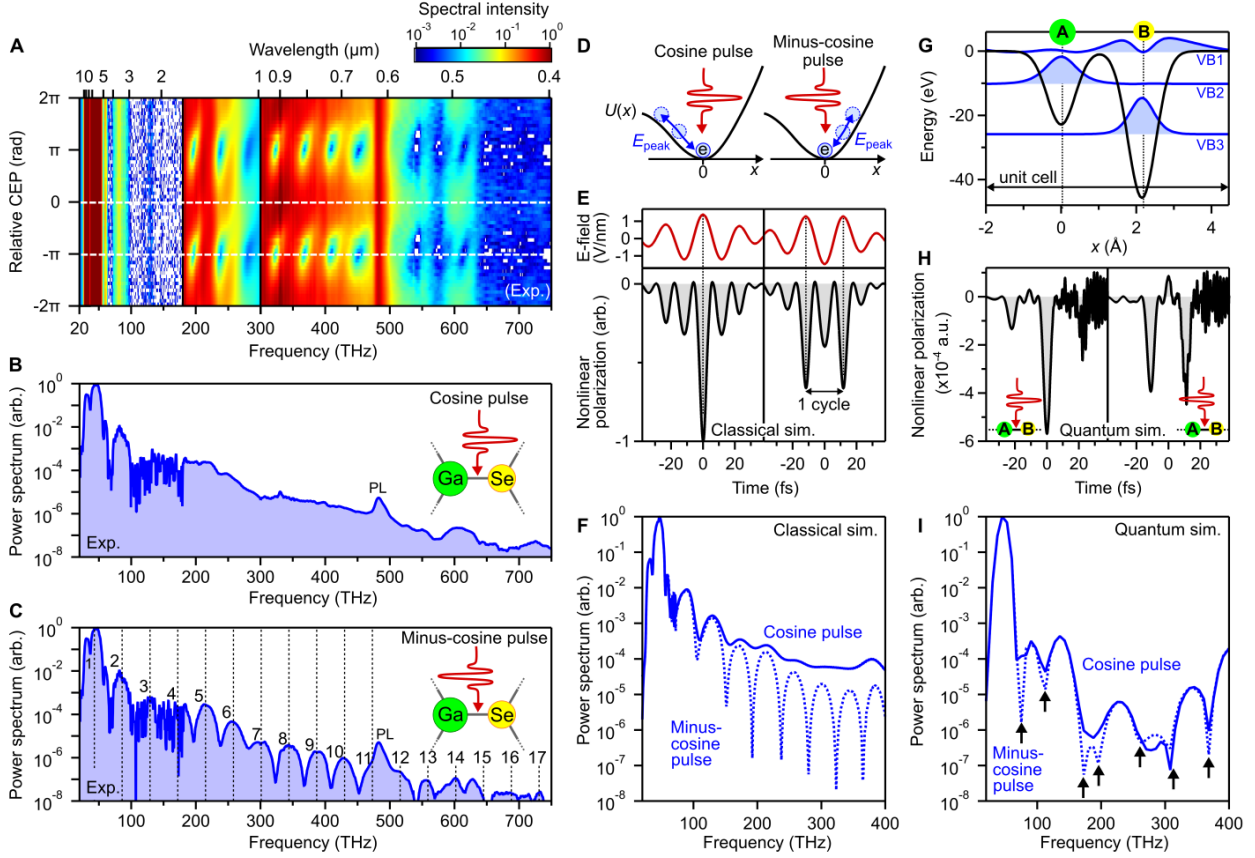

**Fig. 2. Experimental results and real-space explanation.** (**A**) Observed output spectra as a function of the carrier envelope phase (CEP) and polarity of the driving field. (**B**) Experimental output spectrum with cosine excitation (peak field from Ga to Se). (**C**) Experimental output spectrum with minus-cosine excitation (peak field from Se to Ga). (**D**) Concept of classical 1D anharmonic oscillator model. The electron (blue) in an asymmetric potential is driven by a laser field (red). (**E**) Time-dependent nonlinear polarization (black) produced by the laser waveforms (red) as predicted by the classical model. (**F**) Simulated output spectra from the classical simulation for cosine excitation (solid blue curve) and minus-cosine excitation (dashed blue line). (**G**) Model potential for the non-perturbative 1D quantum simulation (black) and electron density of the occupied valence bands (blue). (**H**) Nonlinear polarization from the quantum simulation. Fast oscillations at large times are due to residual nonlinear absorption. (**I**) Simulated output spectra from the quantum-mechanical model for the cosine driving pulse (solid blue curve) and minus-cosine pulse (dashed blue line). The dip positions (arrows) are not exactly between the harmonic orders because the energy of maximum destructive interference is shifted by the unequal intensities of neighbouring harmonics.

Figure 2F shows the results for $\varphi_{CE} = 0$ and $\varphi_{CE} = \pi$. Both simulated spectra of this one-dimensional, classical model reproduce nearly all the observed features in the experiment and in particular the appearance and disappearance of the modulation into harmonic orders when changing from a cosine to a minus-cosine excitation field. In order to elucidate the underlying physics, we plot in Fig. 2E the nonlinear polarization in the time domain. For a cosine-shaped excitation waveform with peak towards the steeper potential gradient (left panel in Fig. 2D), the nonlinear polarization is predominantly induced only once, at the driving laser's highest field peak. Such an isolated burst of nonlinear polarization corresponds in the spectrum to a broadband, featureless emission like observed in the experiment of Fig. 2B. In contrast, a minus-cosine driving pulse (right panel in Fig. 2D) with peak field towards the shallower side of the potential produces substantially weaker nonlinear polarization at the peak time but instead two emission bursts at the times of the two second-most intense field cycles in opposite direction. Spectral interference between these two nonlinear polarization bursts, separated in time by one optical period, explains the measured periodic interferences in the frequency domain (see Fig. 2C).

In an equally simple, but quantum-mechanical model, we consider real-space electron dynamics in a one-dimensional crystal whose unit cell is composed of two different atoms A and B (see Figs. 1A and 1B). This model approximates a line cut through the $\varepsilon$-GaSe crystal along one of the bonds. Atom A has lower electron affinity and corresponds to Ga while atom B has a larger electron affinity and corresponds to Se. We choose a unit cell size of $a = 12.2$ au $= 6.5$ Å and atom positions at $x_A = a/3$ and $x_B = 2a/3$, compare with Fig. 1C. The black line in Fig. 2G shows the real-space potential for this asymmetric system; the depths of the two atomic potentials are chosen to reproduce the direct band gap of $\varepsilon$-GaSe of 2.0 eV (*27*). The corresponding band structure is shown in Fig. S6. Before the interaction with the laser, we populate the three lowest bands (VB1 to VB3) with 6 electrons per unit cell, in order to resemble the electronic structure of three-dimensional $\varepsilon$-GaSe (*28*) where the valence electrons are mostly localized around the nuclei and the electrons of VB1 are mostly on Se (*29*). The corresponding ground-state electron densities are depicted in Fig. 2G in blue. The temporal evolution of the electronic system in the near-single-cycle laser waveform is obtained by solving the time-dependent Schrödinger equation (*26*, *30*).

Figure 2I shows results for $\varphi_{CE} = 0$ and $\pi$. The spectrum for the cosine driving pulse with peak field towards atom B (Se) is smooth (solid line) while the spectrum for the minus-cosine pulse with peak field towards atom A (Ga) has harmonic orders with dips in between (dashed line). The time-dependent nonlinear polarizations (see Fig. 2H) again show one or two main bursts, similar to the classical model (compare Fig 2E). Generally, the optical response of condensed matter to long-wavelength radiation is predominantly determined by electrons in the highest occupied valence band, here VB1. As shown in Fig.

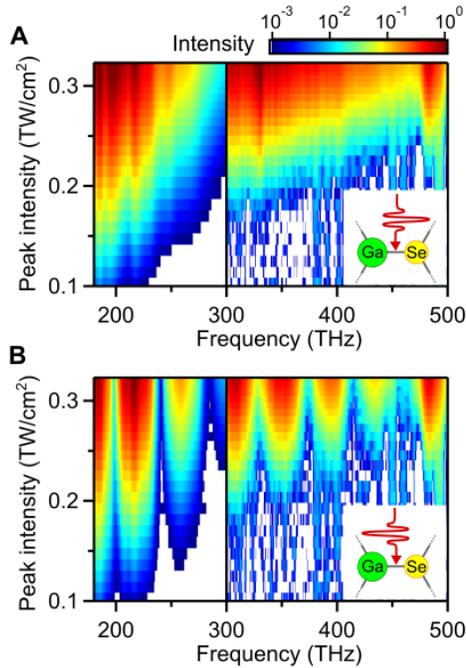

**Fig. 3. Dependence on excitation intensity.** (**A**) Output spectrum for cosine excitation as a function of electric peak field strength. (**B**) Field-strength-dependent output spectrum for minus-cosine excitation. The spectral shape remains the same although the spectral intensities change by more than three orders of magnitude.

2G, the electrons in VB1 are almost entirely localized around atom B, where the potential is asymmetric and less steep towards the direction of the neighbouring atom A. In contrast to the classical model, where the sign of the asymmetry has not been linked to the atoms, the quantum-mechanical model provides this connection and thereby an atomistic explanation of the observed asymmetry results. We conclude that the direction, orientation and different electron affinity of the atoms that form a material's chemical bonds are responsible for the sign, magnitude and shape of the nonlinear-optical response to single-cycle excitation. The atomistic origin of this relation is the quantum-mechanical motion of the bonding electrons in the potential made up for them by the type and distance of the atoms. The time-dependent nonlinear motion of valence electrons is more nonlinear when they are driven towards the shallower side of the potential, towards the neighbouring atom, and weaker towards the steeper side, away from the neighbouring atom.

In order to further verify our conclusions, we report three more experimental investigations. First, we measure spectra at fixed CEP values but for different peak intensities of the driving field. If the field-driven electronic motion along the chemical bonds is dictated by the potential and waveform, the nonlinear output and in particular the absence or appearance of photon-order interferences should not depend on the absolute electric field strength (see also Fig. S5). Figures 3A and 3B show the measured output spectra as a function of excitation intensity for $\varphi_{CE} = 0$ and $\varphi_{CE} = \pi$, respectively. For weaker pump pulses, the harmonic intensities are reduced by up to three-orders of magnitudes, but the general spectral shape with or without harmonic orders remains the same. Second, we report nonlinear emission spectra as a function of crystal-angle with respect to laser polarization ($\theta$ in Fig. 4A). In the experiment, the crystal is rotated around the *c*-axis. $\varepsilon$-GaSe has a three-fold symmetry around the *c* axis (space group $D_3^h$). It repeats itself every 120°, but opposite Ga-Se bonds occur every 60° (see Fig. 4A). We first report angle-dependent results for an excitation waveform that is chirped to a duration of 3.8 optical cycles, in order to elucidate the crystal's response to multi-cycle waveforms (*23*, *31*). Figure 4D shows that the harmonic-order interference appears every 60°, although a rotation of 120° is required to reproduce the original atomic structure. A 3.8-cycle excitation field has too many optical cycles with nearly equal field strength in both directions and therefore averages out the underlying asymmetric response and directivity of the heteronuclear bonds. In other words, Ga-Se and Se-Ga produce the same result. In contrast, Figs. 4B and 4C show the output spectrum as a function of angle for the case of single-cycle excitation fields, again for $\varphi_{CE} = 0$ and $\pi$. We observe two distinct results. First, the 60° periodicity observed with the longer pulses disappears, and instead a 120° periodicity shows up, corresponding to the angle between the three Ga-Se bonds when taking into account the bond direction. Second, when comparing the cosine and minus-cosine results of Figs. 4B and 4C, we see that the whole pattern shifts by 60°. This value reflects the difference of 180° between a cosine and a minus-cosine pulse minus the 120° of the Ga-Se bonds. As a third investigation, we repeated all the above measurements with a 55-µm thick $\varepsilon$-GaSe crystal, twice thicker than the crystal used above. We obtained qualitatively identical results (*26*), demonstrating that all observed phenomena indeed originate from atomic-scale electron dynamics and not from any macroscopic optical pulse propagation or dispersion effects (*32*).

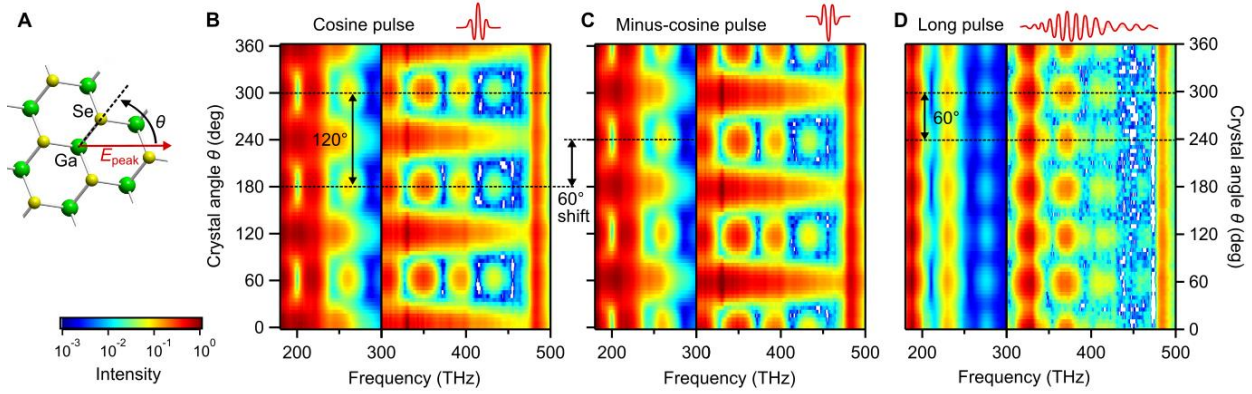

**Fig. 4. Scan of crystal angle and absolute bond orientation.** (**A**) Definition of the crystal angle $\theta$ between the peak electric field vector direction and the Ga-Se bond direction. (**B**) Output spectrum as a function of $\theta$ for cosine excitation. (**C**) Output spectrum for minus-cosine pulse. The periodicity of 120° and the shift to each other by 60° reveal the polar chemical bonds. (**D**) Angle scan with longer pulses (3.8 cycles). A periodicity of 60° shows the averaging of the fundamental asymmetric response by the multiple optical cycles.

In combination with the results of Fig. 2, these observations establish an atomistic picture of nonlinear optics on basis of the chemical bonds. The temporal shape of the nonlinear optical polarization on the level of a single optical excitation cycle is dictated by field-driven valence electronic dynamics on atomic dimensions and for heteronuclear materials by the orientation, angles and polarity of the chemical bonds. The absolute direction of the electric field from Ga to Se is relevant for either the production of a broadband, featureless spectrum that is related to the emission of an isolated optical output pulse (*33*) or a spectrum in form of even and odd harmonics that is related to multiple light emissions via spectral interference (*16*, *23*). A material's chemical structure and in particular the polarity and absolute orientation of the bonds with respect to the incoming light cycles determines the macroscopic outcome of the nonlinear interaction.

**Acknowledgements:** We thank Christina Hofer and Ioachim Pupeza for helpful discussions on mid-infrared optics. Y.M. acknowledges Peter Hommelhoff for his general support.

**Funding:** This work was supported by the European Research Council (CoG No. 647771), the Munich-Centre for Advanced Photonics, CREST (JPMJCR15N1), JSPS KAKENHI No. 18K14145 and 19H00869, Exploratory Challenge on Post-K Computer from the Ministry of Education, Culture, Sports, Science and Technology (MEXT) of Japan, the Center of Innovation Program from the Japan Science and Technology Agency, JST, and MEXT Quantum Leap Flagship Program (MEXT Q-LEAP) Grant Number JPMXS0118067246.

**Author contributions:** Y.M. conceived and performed the experiment. B.C. and Y.M. prepared the light source. Y.S., M.T. and Y.M. performed the quantum mechanical simulations. Y.M. performed the classical simulations. P.B. and K.L.I. supervised the work. Y.M., Y.S. and P.B. wrote the manuscript with input from all authors.

**Competing interests:** The authors declare no competing interests.

**Data and materials availability:** All relevant data are available in the main text or the supplementary materials.

**References and notes**

1. R. W. Boyd, *Nonlinear Optics* (Academic Press, 2008).

2. G. S. He, *Nonlinear Optics and Photonics* (Oxford University Press, 2014).

3. T. T. Luu, M. Garg, S. Yu. Kruchinin, A. Moulet, M. T. Hassan, E. Goulielmakis, Extreme ultraviolet high-harmonic spectroscopy of solids. *Nature*. **521**, 498–502 (2015).

4. M. Lucchini, S. A. Sato, J. Herrmann, A. Ludwig, M. Volkov, L. Kasmi, Y. Shinohara, K. Yabana, L. Gallmann, U. Keller, Observation of femtosecond dynamical franz-keldysh effect in polycrystalline diamond. *Science*. **353**, 916–919 (2016).

5. A. Schiffrin, T. Paasch-Colberg, N. Karpowicz, V. Apalkov, D. Gerster, S. Mühlbrandt, M. Korbman, J. Reichert, M. Schultze, S. Holzner, J. V. Barth, R. Kienberger, R. Ernstorfer, V. S. Yakovlev, M. I. Stockman, F. Krausz, Optical-field-induced current in dielectrics. *Nature*. **493**, 70–74 (2013).

6. A. Moulet, J. B. Bertrand, T. Klostermann, A. Guggenmos, N. Karpowicz, Soft x-ray excitonics. *Science*. **1138**, 1134–1138 (2017).

7. A. Sommer, E. M. Bothschafter, S. A. Sato, C. Jakubeit, T. Latka, O. Razskazovskaya, H. Fattahi, M. Jobst, W. Schweinberger, V. Shirvanyan, V. S. Yakovlev, R. Kienberger, K. Yabana, N. Karpowicz, M. Schultze, F. Krausz, Attosecond nonlinear polarization and light-matter energy transfer in solids. *Nature*. **534**, 86–90 (2016).

8. M. Volkov, S. A. Sato, F. Schlaepfer, L. Kasmi, N. Hartmann, M. Lucchini, L. Gallmann, A. Rubio, U. Keller, Attosecond screening dynamics mediated by electron localization in transition metals. *Nat. Phys.* (2019), doi:10.1038/s41567-019-0602-9.

9. G. Vampa, P. B. Corkum, T. Brabec, Light amplification by seeded Kerr instability. *Science*. **675**, 673–675 (2018).

10. A. H. Chin, O. G. Calderón, J. Kono, Extreme midinfrared nonlinear optics in semiconductors. *Phys. Rev. Lett.* **86**, 3292–3295 (2001).

11. H. A. Hafez, S. Kovalev, J. C. Deinert, Z. Mics, B. Green, N. Awari, M. Chen, S. Germanskiy, U. Lehnert, J. Teichert, Z. Wang, K. J. Tielrooij, Z. Liu, Z. Chen, A. Narita, K. Müllen, M. Bonn, M. Gensch, D. Turchinovich, Extremely efficient terahertz high-harmonic generation in graphene by hot Dirac fermions. *Nature*. **561**, 507–511 (2018).

12. S. Han, L. Ortmann, H. Kim, Y. W. Kim, T. Oka, A. Chacon, B. Doran, M. Ciappina, M. Lewenstein, S.-W. Kim, S. Kim, A. S. Landsman, Extraction of higher-order nonlinear electronic response in solids using high harmonic generation. *Nat. Commun.* **10**, 3272 (2019).

13. S. Ghimire, A. D. Dichiara, E. Sistrunk, P. Agostini, L. F. Dimauro, D. A. Reis, Observation of high-order harmonic generation in a bulk crystal. *Nat. Phys.* **7**, 138–141 (2010).

14. O. Schubert, M. Hohenleutner, F. Langer, B. Urbanek, C. Lange, U. Huttner, D. Golde, T. Meier, M. Kira, S. W. Koch, R. Huber, Sub-cycle control of terahertz high-harmonic generation by dynamical Bloch oscillations. *Nat. Photonics*. **8**, 119–123 (2014).

15. G. Vampa, T. J. Hammond, N. Thiré, B. E. Schmidt, F. Légaré, C. R. McDonald, T. Brabec, P. B. Corkum, Linking high harmonics from gases and solids. *Nature*. **522**, 462–464 (2015).

16. M. Hohenleutner, F. Langer, O. Schubert, M. Knorr, U. Huttner, S. W. Koch, M. Kira, R. Huber, Real-time observation of interfering crystal electrons in high-harmonic generation. *Nature*. **523**, 572–575 (2015).

17. A. A. Lanin, E. A. Stepanov, A. B. Fedotov, A. M. Zheltikov, Mapping the electron band structure by intraband high-harmonic generation in solids. *Optica*. **4**, 516–519 (2017).

18. H. Liang, P. Krogen, Z. Wang, H. Park, T. Kroh, K. Zawilski, P. Schunemann, J. Moses, L. F. DiMauro, F. X. Kärtner, K.-H. Hong, High-energy mid-infrared sub-cycle pulse synthesis from a parametric amplifier. *Nat. Commun.* **8**, 141 (2017).

19. N. Yoshikawa, T. Tamaya, K. Tanaka, High-harmonic generation in graphene enhanced by elliptically polarized light excitation. *Science*. **356**, 736–738 (2017).

20. H. Shirai, F. Kumaki, Y. Nomura, T. Fuji, High-harmonic generation in solids driven by subcycle midinfrared pulses from two-color filamentation. *Opt. Lett*. **43**, 2094 (2018).

21. U. Huttner, M. Kira, S. W. Koch, Ultrahigh Off-Resonant Field Effects in Semiconductors. *Laser Photon. Rev*. **11**, 1700049 (2017).

22. Y. S. You, D. A. Reis, S. Ghimire, Anisotropic high-harmonic generation in bulk crystals. *Nat. Phys.* **13**, 345–349 (2016).

23. F. Langer, M. Hohenleutner, U. Huttner, S. W. Koch, M. Kira, R. Huber, Symmetry-controlled temporal structure of high-harmonic carrier fields from a bulk crystal. *Nat. Photonics*. **11**, 227–231 (2017).

24. B.-H. Chen, E. Wittmann, Y. Morimoto, P. Baum, E. Riedle, Octave-spanning single-cycle middle-infrared generation through optical parametric amplification in LiGaS$_2$. *Opt. Express*. **27**, 21306 (2019).

25. N. Kuroda, O. Ueno, Y. Nishina, Lattice-dynamical and photoelastic properties of GaSe under high pressures studied by Raman scattering and electronic susceptibility. *Phys. Rev. B*. **35**, 3860 (1987).

26. Materials and methods are available as supplementary materials.

27. K. R. Allakhverdiev, M. Ö. Yetis, S. Özbek, T. K. Baykara, E. Y. Salaev, Effective nonlinear GaSe crystal. Optical properties and applications. *Laser Phys.* **19**, 1092–1104 (2009).

28. E. Mooser, I. C. Schlüter, M. Schlüter, The electronic charge densities in semiconducting layer and chain structures. *J. Phys. Chem. Solids*. **35**, 1269–1284 (1974).

29. D. V. Rybkovskiy, A. V. Osadchy, E. D. Obraztsova, Transition from parabolic to ring-shaped valence band maximum in few-layer GaS, GaSe, and InSe. *Phys. Rev. B*. **90**, 235302 (2014).

30. T. Ikemachi, Y. Shinohara, T. Sato, J. Yumoto, M. Kuwata-Gonokami, K. L. Ishikawa, Trajectory analysis of high-order-harmonic generation from periodic crystals. *Phys. Rev. A*. **95**, 043416 (2017).

31. K. Kaneshima, Y. Shinohara, K. Takeuchi, N. Ishii, K. Imasaka, T. Kaji, S. Ashihara, K. L. Ishikawa, J. Itatani, Polarization-Resolved Study of High Harmonics from Bulk Semiconductors. *Phys. Rev. Lett.* **120**, 243903 (2018).

32. P. Xia, C. Kim, F. Lu, T. Kanai, H. Akiyama, J. Itatani, N. Ishii, Nonlinear propagation effects in high harmonic generation in reflection and transmission from gallium arsenide. *Opt. Express*. **26**, 29393–29400 (2018).

33. M. Garg, M. Zhan, T. T. Luu, H. Lakhotia, T. Klostermann, A. Guggenmos, E. Goulielmakis, Multi-petahertz electronic metrology. *Nature*. **538**, 359–363 (2016).